# Ion modes in dense ionized plasmas through nonadiabatic molecular dynamics

R. A. Davis ,[*] W. A. Angermeier, R. K. T. Hermsmeier, and T. G. White[†]

*Physics Department, University of Nevada, Reno, Nevada 89557, USA*

We perform nonadiabatic simulations of warm dense aluminum based on the electron-force field (EFF) variant of wave-packet molecular dynamics. Comparison of the static ion-ion structure factor with density functional theory (DFT) is used to validate the technique across a range of temperatures and densities spanning the warm dense matter regime. Focusing on a specific temperature and density (3.5 eV, 5.2 g/cm$^3$), we report on differences in the dynamic structure factor and dispersion relation across a variety of adiabatic and nonadiabatic techniques. We find the dispersion relation produced with EFF is in close agreement with the more robust and adiabatic Kohn-Sham DFT.

The warm dense matter (WDM) regime defines a dense plasma state where strongly coupled classical ions coexist with partially or fully degenerate electrons [1]. It is a complex state of matter where multibody particle correlations and quantum effects play an important role in determining the overall structure and equation of state [2]. A complete description of WDM is important for describing many physical phenomena, ranging from phase transitions within the interior of large astrophysical objects [3] to temperature relaxation rates during the internal processes of inertial confinement fusion [4].

Strong ion-ion coupling combined with the quantum behavior of the electron fluid make simulation and modeling challenging. One approach that is able to fully capture the strongly coupled behavior of the ions has been molecular dynamics (MD), i.e., a fully atomistic simulation of the ion dynamics that results from numerical integration of the equations of motion [5]. As MD simulations explicitly calculate the ion trajectories they are able to provide transport properties, such as viscosity and thermal diffusivity [6], acoustic properties, such as the sound speed [7,8], and thermodynamic variables, including the equation of state [5,7]. The accuracy and predictive power of such simulations is encompassed within the calculation of the interatomic potential, with increased accuracy coming at the expense of computational cost. Simulations using analytic potentials are able to model the largest systems [9]; however, potentials calculated with orbital-free [7] (OF) or Kohn-Sham [10,11] (KS) density functional theory (DFT) have had the most success when compared to available experimental results [12,13].

MD simulations of dense plasmas typically employ the Born-Oppenheimer (BO) approximation, also known as the adiabatic approximation, usually justified through the disparate energy scales of the electron and ion motion [14]. This approximation is the cornerstone of almost all classical [9,15] and *ab initio* [7,10,11] atomistic simulations; the electrons are assumed to instantaneously adjust to the ion fields, while the ions themselves are confined to a single adiabatic surface [16]. The BO approximation thus neglects important details of the interaction and, of particular relevance to high energy density matter, the dynamics of electron-ion collisions are ignored [17]. Although justified for many equilibrium properties, like the equation of state [8,18], the adiabatic treatment prohibits direct energy transfer between electrons and ions and is therefore problematic for the modeling of transport properties and nonequilibrium matter [17,19,20].

Recent work has investigated the applicability of a Langevin thermostat to capture the effect of electron-ion interactions on the ion dynamics [8,18,21]. In this phenomenological approach, an additional stochastic Gaussian force is added to the equations of motion with a single collision frequency, or friction parameter, that must be determined *a priori* [20]. A number of classical (Rayleigh scattering [22]) and quantum (Born approximation [23]) models exist, but their applicability in the WDM state is unknown. The inclusion of these collisions has been shown to have a profound effect on the ion dynamics [8], including a strongly decreased adiabatic sound speed and increased diffusive mode around $\omega = 0$.

In order to explicitly go beyond the BO approximation, the system must be treated on an electronic time scale while correctly modeling the quantum nature of the electrons. In quantum chemistry, time-dependent DFT is one approach that attempts this. However, for systems made up of the thousands of ions necessary to correctly model the low-frequency ion dynamics, this method remains too computationally intensive [24]. More recently, a technique based on the Bohm theory of quantum mechanics has been implemented. This method propagates an array of thermally coupled $N$ trajectories interacting with a thermally averaged, linearized Bohm potential [16]. An adiabatic sound speed in agreement with the Langevin approach was found, with a collision frequency between the quantum and classical limits. Interestingly, the dynamics were found to have a significantly reduced diffusive mode when compared to the Langevin technique.

Here, we utilize an alternative method, wave packet molecular dynamics (WPMD). It is a time-dependent quantum mechanical technique able to simultaneously simulate (1) the propagation of the ions as classical point particles, and (2) the electrons as quantum mechanical entities [25–29]. The direct inclusion of electrons, and thus the effects of electron-ion interactions, means that WPMD intrinsically goes beyond the adiabatic approximation and is capable of calculating observables in quantum many-body systems [26].

In WPMD, each electron is represented as a wave packet, a spatially localized complex function. It is common to represent a single electron wave packet using a basis set constructed from a sum of $M$ Gaussians. i.e.,

$$\phi(\mathbf{x}) = n^{-1/2} \sum_{\alpha=1}^{M} c_\alpha \varphi_\alpha(\mathbf{x}), \tag{1}$$

$$\varphi_\alpha(\mathbf{x}) = \left(\frac{3}{2\pi s_\alpha^2}\right)^{3/4} e^{-\left(\frac{3}{4s_\alpha^2} - \frac{ip_{s_\alpha}}{2\hbar s_\alpha}\right)(\mathbf{x}-\mathbf{r}_\alpha)^2 + \frac{i}{\hbar}\mathbf{p}_\alpha(\mathbf{x}-\mathbf{r}_\alpha)}, \tag{2}$$

where $n = \sum_{\alpha,\beta} c_\alpha^* c_\beta \int \varphi_\alpha^* \varphi_\beta d^3 x$ is a normalization factor. The remaining terms are dynamical parameters: complex amplitude ($c_\alpha$), position ($\mathbf{r}_\alpha$), momentum ($\mathbf{p}_\alpha$), Gaussian width ($s_\alpha$), and conjugate width momentum ($p_{s_\alpha}$) [25]. These wave packets uniquely define the quantum state of a single electron, with the total many-body wave function constructed from either a Hartree product or Slater determinant [26]. Equations of motion for the dynamical parameters are easily derived from variation of the time-dependent Schrodinger equation [30,31]. One of the great accomplishments of WPMD as a method of modeling plasmas is maintaining the same computational efficiency as in many classical methods [25,27,32].

We utilize a simple variation of WPMD known as electron-force field (EFF). In EFF, electrons are described by a single, floating Gaussian wave packet combined into a many-body wave function using a Hartree product [33,34]. In this case, the equations of motion take on a simple Hamilton form [35],

$$\dot{\mathbf{p}} = -\vec{\nabla} V, \quad \mathbf{p} = m_e \dot{\mathbf{r}}, \quad \dot{p}_s = -\partial V/\partial s, \quad p_s = (3m_e/4),$$

where $V$ is the potential energy of the system and $m_e$ is the electron rest mass. The total energy ($E$) of the system is expressed as the sum of single particle kinetic and electrostatic energies,

$$E = E_{ke} + E_{nuc \cdot nuc} + E_{nuc \cdot elec} + E_{elec \cdot elec} + E_{xc}. \tag{3}$$

For the single Gaussian basis, each term may be written in a simple analytic form,

$$E_{ke} = \frac{\hbar^2}{m_e} \sum_i \frac{3}{2} \frac{1}{s_i^2}, \tag{4}$$

$$E_{nuc(i) \cdot nuc(j)} = \frac{1}{4\pi\epsilon_0} \sum_{i<j} \frac{Z_i Z_j}{R_{ij}}, \tag{5}$$

$$E_{nuc(i) \cdot elec(j)} = -\frac{1}{4\pi\epsilon_0} \sum_{i,j} \frac{Z_i}{R_{ij}} \text{erf}\left(\frac{\sqrt{2}R_{ij}}{s_j}\right), \tag{6}$$

$$E_{elec(i) \cdot elec(j)} = \frac{1}{4\pi\epsilon_0} \sum_{i<j} \frac{1}{R_{ij}} \text{erf}\left(\frac{\sqrt{2}R_{ij}}{\sqrt{s_i^2 + s_j^2}}\right), \tag{7}$$

where $Z$ is the ionic charge, $R$ is the inter-particle separation, and the sum over $i$ and $j$ runs over either all nuclei or electrons. An additional pairwise energy term is introduced between same-spin electrons to account for the lack of antisymmetrization in the Hartree product, while a small repulsive term is added between electrons of opposite spin to prevent electron coalescence,

$$E_{xc} = \sum_{\substack{i,j \\ \text{same spin}}} E(\uparrow\uparrow)_{ij} + \sum_{\substack{i,j \\ \text{opposite spin}}} E(\uparrow\downarrow)_{ij}. \tag{8}$$

The pairwise energy between like and unlike spin electrons is constructed from the Slater and generalized valence bond (GVB) many-electron wave functions and is given by

$$E(\uparrow\uparrow) = E_u - (1+\rho)E_g \quad E(\uparrow\downarrow) = \rho E_g, \tag{9}$$

where $E_u$ is solely the kinetic energy penalty upon pairwise antisymmetrization,

$$E_u = \langle\Psi_S| -\tfrac{1}{2}\nabla^2|\Psi_S\rangle - \langle\Psi_H| -\tfrac{1}{2}\nabla^2|\Psi_H\rangle, \tag{10}$$

and $E_g$ is the energy reduction when using a GVB wave function,

$$E_g = \langle\Psi_{GVB}| -\tfrac{1}{2}\nabla^2|\Psi_{GVB}\rangle - \langle\Psi_H| -\tfrac{1}{2}\nabla^2|\Psi_H\rangle. \tag{11}$$

The many electron wave functions (Slater, GVB, and Hartree) take on their usual pairwise definitions, $\Psi_S \propto (\phi_1(r_1)\phi_2(r_2) - \phi_2(r_1)\phi_1(r_2))$, $\Psi_{GVB} \propto (\phi_1(r_1)\phi_2(r_2) + \phi_2(r_1)\phi_1(r_2))$, and $\Psi_H \propto \phi_1(r_1)\phi_2(r_2)$.

Simple analytic expressions for $E_u$ and $E_g$ may be derived for the single Gaussian basis allowing for rapid evaluation of these expressions. The mixing parameter, $\rho = -0.2$, along with a number of other scaling parameters in the analytic expressions, are matched to a set of molecular test structures [35]. Together, these partially account for the missing electron-ion and electron-electron exchange terms [36]. For many systems, EFF is found to have an exchange energy comparable to unrestricted Hartree-Fock [35]. There have been some recent efforts to include more robust correlation functionals derived from DFT in WPMD [37], but correlation is not currently treated in EFF.

In addition, for high-$Z$ systems, a fixed size wave packet is attached to the otherwise classical ions to represent the strongly bound core electrons. These effective core potentials (ECPs) provide not only an increased computation speed, as the high-frequency oscillations associated with these electrons do not need to be captured, but also increased accuracy as they may exhibit non-Gaussian shapes [36,38]. Due to the strong binding energy and high-frequency oscillations of these core electrons we do not expect them to play a large role in the low-frequency ion dynamics.

EFF is able to simulate systems consisting of thousands of particles for many picoseconds. It has been applied to the study of material properties in extreme environments [34,36,39–41], temperature relaxation rates in warm dense hydrogen [42], and wave-packet spreading in electron-nuclear scattering [25]. However, the low frequency ion modes in dense plasmas have not yet been examined.

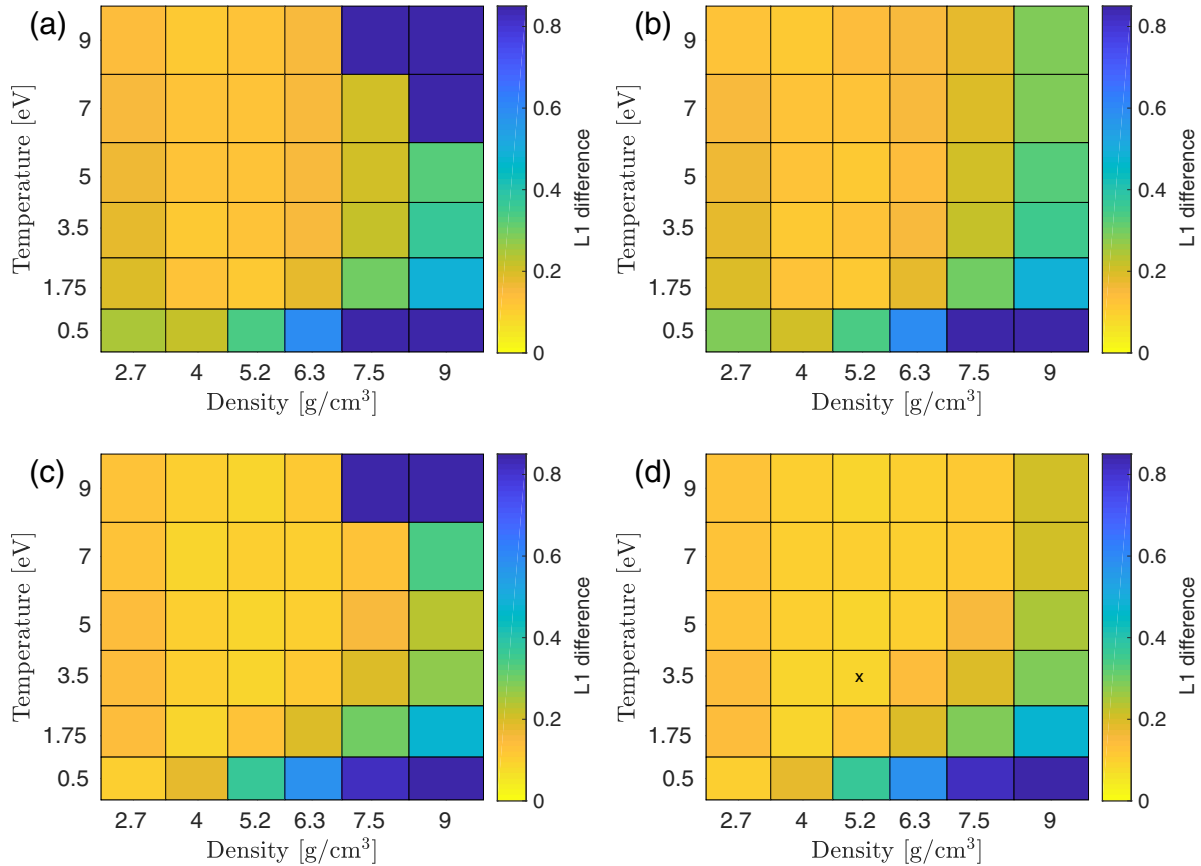

FIG. 1. The L1 difference in the static structure factors for warm dense aluminum are calculated between the electron-force field (EFF) technique, using modified and unmodified parameters, and orbital-free density functional theory (OF-DFT). In each case, lower values for L1 highlight a region of validity for EFF in the low-density/high-temperature regime. (a) and (b) Simulated with the default ECP radius of 1.660 $a_B^{-1}$, whereas (c) and (d) use the improved value of 1.285 $a_B^{-1}$. In (a) and (c) the Gaussians are permitted to expand indefinitely, whereas (b) and (d) are run with the Gaussian width constrained by an additional harmonic potential. The cross in (d), at 3.5 eV and 5.2 g/cm³, marks the best agreement between the two techniques. The SSFs that were used to produce these plots are shown in Figs. S1–S4 in the Supplemental Material [53].

The ion modes are typically investigated through the spatiotemporal Fourier transform of the density-density autocorrelation function, or the dynamic structure factor (DSF). This common metric, used to compare different theoretical techniques, is related to the dynamic and thermodynamic properties of the plasma, but also accessible experimentally, allowing for critical benchmarking of models [12,43]. In particular, collective ion modes are a prominent feature, whose spectra allow for validation and comparison of theoretical predictions for dense matter [44]. For a system in thermodynamic equilibrium, the DSF gives its response to fluctuations of frequency $\omega$ and wave vector $\mathbf{k}$, and is defined as

$$S(\mathbf{k}, \omega) = \frac{1}{2\pi N} \int e^{i\omega t} \langle \rho(\mathbf{k}, t)\rho(-\mathbf{k}, 0)\rangle dt, \quad (12)$$

where $N$ is the total number of particles, and $\langle \dots \rangle$ refers to an ensemble average. Here, $\rho$ is the Fourier transform of the real space time-dependent density distribution. For a system of particles with coordinates $\mathbf{r}_j(t)$ this is given by

$$\rho(\mathbf{k}, t) = \sum_{j=0}^{N} \exp[i\mathbf{k} \cdot \mathbf{r}_j(t)]. \quad (13)$$

The static structure factor (SSF) is the frequency integrated version of this quantity that relates purely to the structure of the system.

To validate the EFF technique across the entire WDM regime, we ran multiple simulations that spanned a range of thermodynamic conditions. The SSF for each condition was calculated using both EFF and OF-DFT, with the L1 difference between the two plotted in Fig. 1(a). Such results can be used to validate EFF throughout the WDM regime, as adiabatic methods have been shown to accurately model the SSF [8]. Our results suggest an area of validity in the low-density region of phase space.

All EFF simulations were performed using the LAMMPS open-source software [45]. Standard periodic boundary conditions were used with $N_i = 216$ aluminum ions and $N_e = 648$ electrons. These simulations were run with a timestep of 1 attosecond for a total of 20 million timesteps, giving an overall simulation time of 20 ps. The inner core electrons were modeled using an effective core potential with a radius of either $1.285\ a_B^{-1}$ or $1.660\ a_B^{-1}$. Before the calculation of any structure factors, the simulation was equilibrated for 90 ps using a Langevin thermostat until the electrons and ions were in equilibrium. After equilibration, each simulation took 9 h on 32 CPUs.

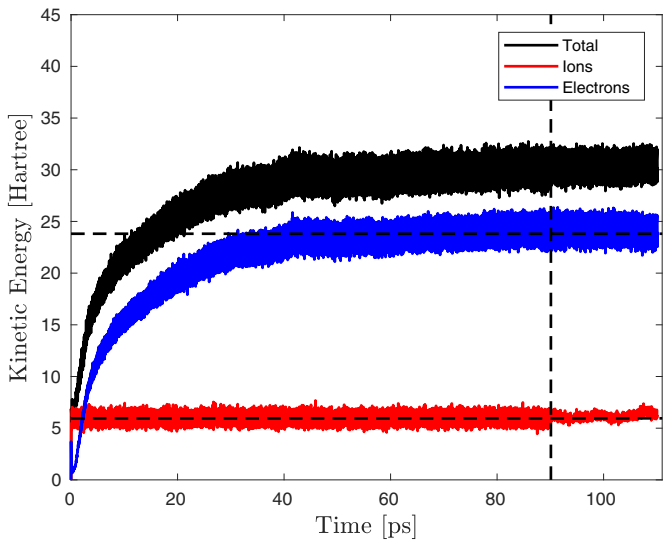

FIG. 2. Temporal evolution of the electronic, ionic, and total kinetic energy in the simulation. Equilibration is achieved over 90 ps with a Langevin thermostat coupled only to the ions. The electrons then equilibrate through electron-electron collisions. At 90 ps, the Langevin thermostat is turned off, and the system is set to run for another 20 ps, during which the structure factors are calculated. The horizontal lines correspond to $\frac{3}{2}N_iT = 5.95$ Ha and $\frac{4}{2}N_eT = 23.8$ Ha, respectively.

Temperature equilibration in the simulation employed a Langevin thermostat coupled only to the ions, with the temperature of the electron subsystem allowed to equilibrate purely through electron-ion collisions. The built-in EFF temperature control, which couples to both electrons and ions, was found to lead to nonequilibrium velocity distributions. Figure 2 shows the evolution of the energy in each subsystem throughout the $T$=0.5 eV simulation, with similar behavior found at all temperatures and densities. At 90 ps, the thermostat is removed, and the static and dynamic structure factors calculated. It should be noted that the electron energy approaches $\frac{4}{2}N_eT = 23.8$ Ha and the ion energy $\frac{3}{2}N_iT = 5.95$ Ha. In the WPMD approximation, the Gaussian width is taken to be a classical fourth degree of freedom for each electron [25].

The OF-DFT simulations used to generate Fig. 1 were performed in the open-source software ABINIT [46,47] using the Thomas-Fermi module and the exchange-correlation terms in local density approximation of Perdew-Zunger-Ceperley-Alder [48]. For the interaction between ions and electrons, a bulk-local pseudopotential [49] is used in which the 10 inner electrons are frozen, while the three valence electrons are explicitly treated. Simulations employed a 216-ion cubic supercell with periodic boundary conditions. They ran for 5 ps and used a 0.5-fs timestep. A plane-wave energy cutoff of 30 eV was used with 880 bands. To control the temperature, a Nose-Hoover thermostat [50] with an inertia parameter corresponding to a temperature oscillation period of over 100 timesteps was used. This ensures weak coupling to the heat bath and a correctly sampled canonical distribution. After equilibration, each OF-DFT simulation took 20 h on 32 CPUs.

Two modifications to the EFF technique were required to improve agreement with the OF-DFT simulations for warm dense aluminum. The first, as also found in Refs. [27,42,51], is the inclusion of a harmonic potential to constrain the width of the Gaussian wave packet. In EFF, the restraining harmonic potential takes the form $E = \frac{1}{2}k_s s^2$ for $s > L_{box/2}$, where $k_s = 1$ Hartrees/Bohr$^2$. Such a constraint is needed to prevent the size of the wave packet from increasing without limit and leading to unphysical diminished electron-ion and electron-electron interactions. While such expansion is not itself unphysical, a wave packet will naturally spread when not confined by a potential; the limited basis set prevents localization around ions. It is this lack of localization that leads to unphysically low interaction. The improvement afforded by including a harmonic potential is shown in Fig. 1(b), where much better agreement is seen in the high-temperature/high-density region.

The second change made was to alter the width of the ECP Gaussian, the Gaussian attached to the ions representing the 10 inner shell electrons. For aluminum, EFF uses a radius of $1.66\ a_B^{-1}$, determined from the ground state of face-centered cubic bulk aluminum [38]. We found that this number needed to be decreased to $1.285\ a_B^{-1}$, much closer to the radius of an Al$^{3+}$ ion [52]. Figure 1(c) demonstrates the improvement afforded by this change, while Fig. 1(d) implements both modifications for the best agreement. We refer to EFF run with both modifications as modified EFF (mEFF). All of the SSFs that were used to produce these plots are provided in Figs. S1–S4 of the Supplemental Material [53].

The biggest disagreement between mEFF and OF-DFT is found in the high-density/low-temperature regime; this difference is attributed to crystallization in the OF-DFT simulations that is not apparent with EFF. However, with neither KS-DFT simulations or experimental results available in this regime, the validity of both techniques may be questioned. For four thermodynamic conditions, we were able to compare the OF-DFT/mEFF/EFF results to published KS-DFT SSFs [7,8,12,54]. Close agreement was found in three cases: 2.7 g/cm$^3$ and 0.5 eV, 2.7 g/cm$^3$ and 5 eV, and 5.2 g/cm$^3$ and 3.5 eV. As shown in Fig. 3(a), the results at 6.3 g/cm$^3$ and 1.75 eV disagreed. At this condition, mEFF underestimates the degree of ion-ion correlation in the system.

The cross in Fig. 1(d), at 3.5 eV and 5.2 g/cm$^3$, marks the best agreement between the two techniques. Figure 3(b) shows the SSF calculated with mEFF and EFF at this condition. This thermodynamic condition allows for direct comparison with a number of previously published results, both adiabatic and nonadiabatic, and the SSFs calculated with the more computationally intensive OF-DFT and KS-DFT MD are included. Excellent agreement is found between the SSF calculated with mEFF and with OF-DFT. Both results take similar computational times; however, mEFF explic-itly includes the electron dynamics. The KS-DFT results are currently the most accurate available, but extremely computationally intensive. They exhibit an increased peak height around $k = 2\,a_B^{-1}$, suggesting that even at this condition, neither EFF or OF-DFT completely captures the ionic structure in the system.

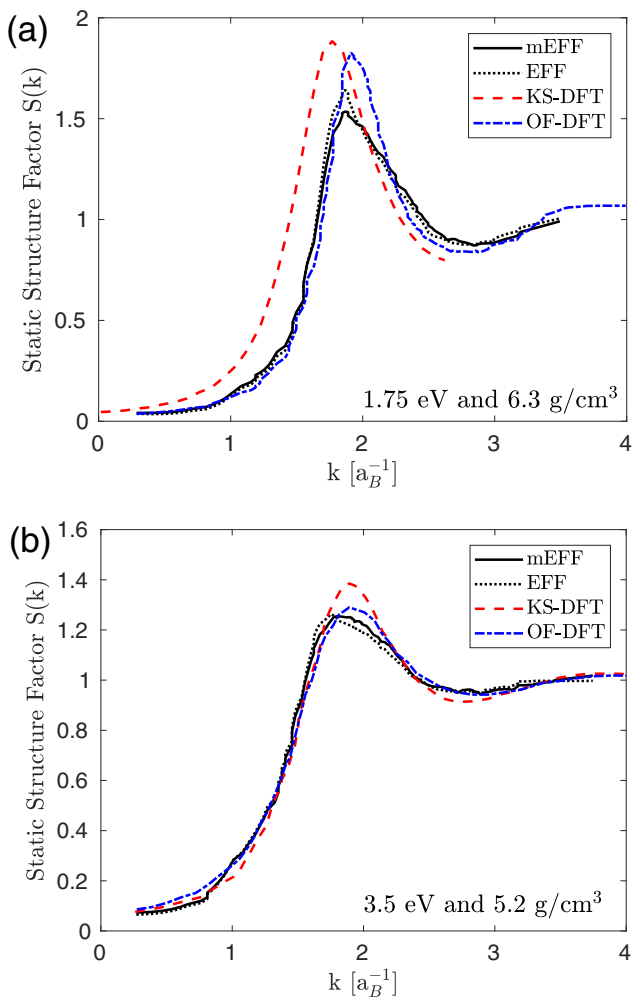

FIG. 3. The static structure factor for warm dense aluminum calculated with the nonadiabatic electron-force field method; both with its modified (mEFF) and original (EFF) parameters. (a) 1.75 eV and 6.4 g/cm$^3$; (b) 3.5 eV and 5.2 g/cm$^3$. For comparison, results from the more computationally intensive orbital-free density functional theory (OF-DFT) and Kohn-Sham density functional theory (KS-DFT) are included [7,8,12,54].

We now turn our attention to the dynamic plasma properties, which are expected to vary more with the inclusion of nonadiabatic effects. We focus our attention on the thermodynamic condition where we find best agreement in the SSF between all techniques, at 5.2 g/cm$^3$ and 3.5 eV. Figure 4(a) shows the DSF calculated with mEFF and KS-DFT. The mEFF results are well matched by the KS-DFT results, agreeing at the highest and lowest $k$ values tested; small differences are apparent in the shape of the spectra at the intermediate spatial scale (i.e., $k = 0.76\,a_B^{-1}$). At this condition, as with the SSF, both EFF and mEFF exhibit near identical features [see Fig. 4(b)]. Interestingly, as with the Bohmian mechanics

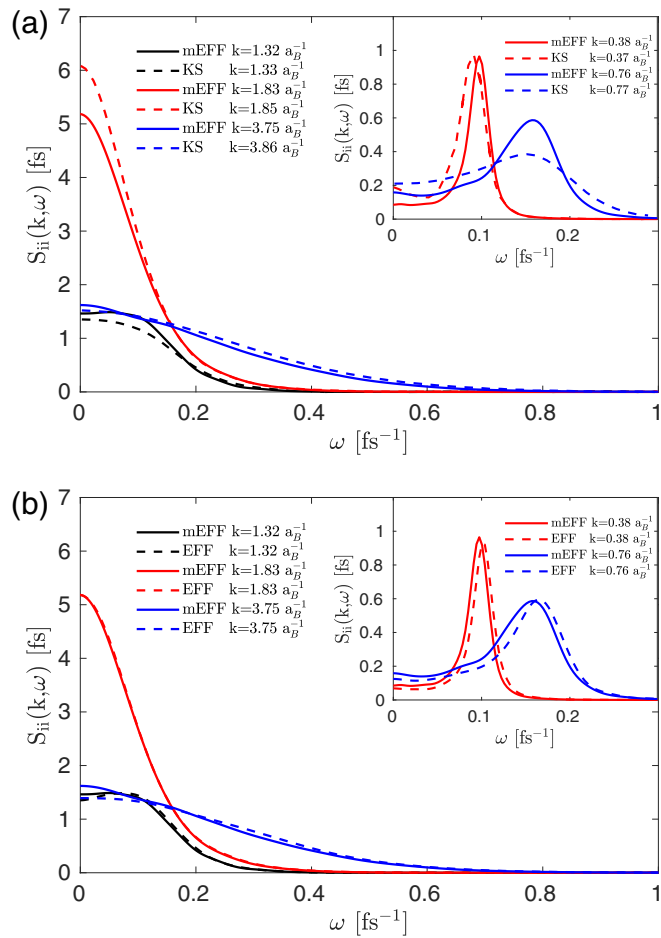

FIG. 4. The dynamic structure factor for warm dense aluminum at 3.5 eV and 5.2 g/cm$^3$ calculated with the nonadiabatic electron-force field method, both with its improved (mEFF) and original (EFF) parameters. For comparison, results from Kohn-Sham density functional theory (KS-DFT) are included [10].

technique, we see a strongly reduced diffusive mode in the DSF region $\omega = 0$.

The peak position is excellently matched across all $k$ values, as exemplified in the dispersion relation shown in Fig. 5. We find excellent agreement between the mEFF method and the KS-DFT results. Both techniques agree on the adiabatic sound speed (given by the low-$k$ slope of the dispersion relation) as well as on their prediction for the onset and strength of negative dispersion. For comparison, we have included in Fig. 5 the results from OF-DFT, with a Langevin thermostat, and those calculated from the Bohmian mechanics technique. While the Bohmian mechanics results are closer to those with a higher collision frequency, highlighting the importance of nonadiabatic techniques, the EFF results suggest that the electrons may have less of an effect on the ion dynamics than previously thought.

Despite its success, the comparison with KS-DFT suggests that mEFF breaks down at the extremum of temperature and density. Even at our trial condition of 3.5 eV and 5.2 g/cm$^3$, there is a lack of correlation in the SSF when compared to KS-DFT. Furthermore, the discrepancy in shape of the DSF at $k = 0.76\,a_B^{-1}$ between mEFF and KS-DFT suggests there may be differences in transport properties between the two models. To address these discrepancies, we suggest improvements to the mEFF formalism that tackle three fundamental approximations. These are (1) a limited basis set consisting of a single Gaussian wave-packet; (2) neglect of the electron-electron and electron-ion components of the exchange energy; and (3) neglect of correlation energy. The first of these limits wave-packet breakup and has previously been identified as a flaw in WPMD [25]. Work to expand the region of applicability of mEFF should focus on these three limitations.

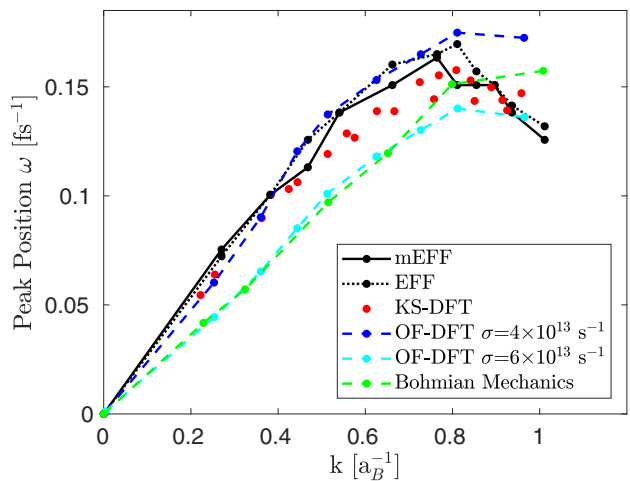

FIG. 5. The dispersion relation for warm dense aluminum at 3.5 eV and 5.2 g/cm³, calculated with the nonadiabatic electron-force field method, both with its modified (mEFF) and original (EFF) parameters. For comparison, results from Kohn-Sham density functional theory (KS-DFT) [10], orbital-free density functional theory (OF-DFT) with a Langevin thermostat [8] at two friction parameters, and the result of Bohmian mechanics [16] are included.

In summary, the mEFF technique demonstrates reasonable agreement to the static properties of warm dense aluminum, obtained with the more robust density functional theory, across a large portion of the WDM phase space, with the best agreement found in the low-density/high-temperature regime. This is achieved with less computational effort and full treatment of the electron dynamics. The effect of electron-ion correlations on the ion modes were investigated and found to have less effect than when utilizing Bohmian mechanics [16], with a dispersion relation in closer agreement with the adiabatic KS-DFT results. These results directly address concerns over WPMD, as an *ab initio* many-body method for dense plasmas [26], providing corroboration with other models.

Overall, there is still much we do not understand about the ion dynamics in warm dense matter. The results of the Langevin thermostat, Bohmian mechanics technique, and now WPMD techniques play an important role in our understanding of the ion modes, sound speeds, and transport coefficients in dense plasmas. Of course, experimental verification along with further computational investigation is essential.

**ACKNOWLEDGMENTS**

This material is partially based upon work supported by the U.S. Department of Energy, Office of Science, Office of Fusion Energy Sciences under Grant No. DE-SC0019268. The authors acknowledge the support of Research & Innovation and the Office of Information Technology at the University of Nevada, Reno for computing time on the Pronghorn High-Performance Computing Cluster.

[1] S. Ichimaru, Rev. Mod. Phys. **54**, 1017 (1982).

[2] F. Grazani, M. P. Desjarlais, R. Redmer, and S. B. Trickey, *Frontiers and Challenges in Warm Dense Matter* (Springer, Berlin, 2014).

[3] T. Guillot, Science **286**, 72 (1999).

[4] J. Lindl, Phys. Plasmas **2**, 3933 (1995).

[5] F. Graziani *et al.*, High Energy Density Phys. **8**, 105 (2012).

[6] G. Salin and J. Caillol, Phys. Plasmas **10**, 1220 (2003).

[7] T. G. White, S. Richardson, B. J. B. Crowley, L. K. Pattison, J. W. O. Harris, and G. Gregori, Phys. Rev. Lett. **111**, 175002 (2013).

[8] P. Mabey *et al.*, Nat. Commun. **8**, 14125 (2017).

[9] J. Vorberger, Z. Donko, I. M. Tkachenko, and D. O. Gericke, Phys. Rev. Lett. **109**, 225001 (2012).

[10] H. R. Rüter and R. Redmer, Phys. Rev. Lett. **112**, 145007 (2014).

[11] B. B. L. Witte, L. B. Fletcher, E. Galtier, E. Gamboa, H. J. Lee, U. Zastrau, R. Redmer, S. H. Glenzer, and P. Sperling, Phys. Rev. Lett. **118**, 225001 (2017).

[12] L. B. Fletcher *et al.*, Nat. Phot. **9**, 274 (2015).

[13] T. Ma *et al.*, Phys. Rev. Lett. **110**, 065001 (2013).

[14] M. Born and J. R. Oppenheimer, Ann. Phys. **84**, 457 (1927).

[15] J. P. Mithen, J. Daligault, B. J. B. Crowley, and G. Gregori, Phys. Rev. E **84**, 046401 (2011).

[16] B. Larder *et al.*, Science Advances **5**, eaaw1634 (2019).

[17] J. Clérouin, G. Robert, P. Arnault, C. Ticknor, J. D. Kress, and L. A. Collins, Phys. Rev. E **91**, 011101(R) (2015).

[18] J. Dai and J. Yuan, Europhysics Letters Association **88**, 20001 (2009).

[19] T. G. White *et al.*, Phys. Rev. Lett. **112**, 145005 (2014).

[20] Zh. A. Moldabekov, T. Dornheim, M. Bonitz, and T. S. Ramazanov, Phys. Rev. E **101**, 053203 (2020).

[21] A. Tamm, M. Caro, A. Caro, G. Samolyuk, M. Klintenberg, and A. A. Correa, Phys. Rev. Lett. **120**, 185501 (2018).

[22] A. V. Plyukhin, Phys. Rev. E **77**, 061136 (2008).

[23] R. Redmer *et al.*, Plasma Sci. IEEE Trans. **33**, 77 (2005).

[24] X. Li *et al.*, J. Chem. Phys. **123**, 084106 (2005).

[25] P. E. Grabowski, A. Markmann, I. V. Morozov, I. A. Valuev, C. A. Fichtl, D. F. Richards, V. S. Batista, F. R. Graziani, M. S. Murillo *et al.*, Phys. Rev. E **87**, 063104 (2013).

[26] P. E. Grabowski, in *Frontiers and Challenges in Warm Dense Matter*, edited by F. Grazani, M. P. Desjarlais, R. Redmer, and S. B. Trickey (Springer, Berlin, 2014), pp. 265–282.

[27] I. V. Morozov and I. A. Valuev, J. Phys. A **42**, 214044 (2009).

[28] I. V. Morozov and I. A. Valuev, Contrib. Plasma Phys. **52**, 140 (2012).

[29] I. A. Valuev and I. V. Morozov, J. Phys.: Conf. Ser. **653**, 012153 (2015).

[30] H. Feldmeier and J. Schnack, Rev. Mod. Phys. **72**, 655 (2000).

[31] D. Klakow, C. Toepffer, and P. G. Reinhard, J. Chem. Phys. **101**, 10766 (1994).

[32] M. P. Allen and D. J. Tildesley, *Computer Simulation of Liquids* (Oxford University Press, Oxford, 2017).

[33] J. T. Su and W. A. Goddard, Phys. Rev. Lett. **99**, 185003 (2007).

[34] J. T. Su and W. A. Goddard, J. Chem. Phys. **131**, 244501 (2009).

[35] J. T. Su and W. Goddard, Proc. Natl. Acad. Sci. USA **106**, 1001 (2009).

[36] H. Xiao, A. Jaramillo-Botero, and W. A. Goddard, Mech. Mater. **90**, 243 (2015).

[37] Y. S. Lavrinenko, I. V. Morozov, and I. A. Valuev, Contrib. Plasma Phys. **59**, e201800179 (2019).

[38] H. Xiao, Ph.D thesis, California Institute of Technology, Pasadena, 2015.

[39] A. Jaramillo-Botero, J. Su, A. Qi, and W. A. Goddard, J. Comput. Chem. **32**, 497 (2011).

[40] H. Kim, J. T. Su, and W. A. Goddard, Proc. Natl. Acad. Sci. USA **108**, 15101 (2011).

[41] M. Lan, Z. Yang, and X. Wang, Comput. Mater. Sci. **179**, 109697 (2020).

[42] Q. Ma, J. Dai, D. Kang, M. S. Murillo, Y. Hou, Z. Zhao, J. Yuan *et al.*, Phys. Rev. Lett. **122**, 015001 (2019).

[43] S. H. Glenzer and R. Redmer, Rev. Mod. Phys. **81**, 1625 (2009).

[44] G. Gregori and D. O. Gericke, Phys. Plasmas **16**, 056306 (2009).

[45] S. J. Plimpton, J. Comput. Phys. **117**, 1 (1995).

[46] X. Gonze *et al.*, Comput. Phys. Comm. **180**, 2582 (2009).

[47] X. Gonze *et al.*, Comput. Phys. Comm. **205**, 106 (2016).

[48] J. P. Perdew and A. Zunger, Phys. Rev. B **23**, 5048 (1981).

[49] C. Huang and E. A. Carter, Phys. Chem. Chem. Phys. **10**, 7109 (2008).

[50] S. Nosé, Mol. Phys. **52**, 255 (1984).

[51] M. Knaup, P.-G. Reinhard, C. Toepffer, and G. Zwicknagel, Comput. Phys. Commun. **147**, 202 (2002).

[52] R. D. Shannon, Acta Crystallogr. A **32**, 751 (1976).

[53] See Supplemental Material at http://link.aps.org/supplemental/10.1103/PhysRevResearch.2.043139 for all SSF's produced by EFF and OFDFT to generate Fig. 1.

[54] C. E. Starrett and D. Saumon, Phys. Rev. E **92**, 033101 (2015).